\documentclass[usegraphicx,usenatbib]{mn2e}

\usepackage{times}
\usepackage{amsmath}
\usepackage{amssymb}

\begin{document}

\title[Eddington ratio and accretion efficiency in AGN evolution]{Eddington ratio and accretion efficiency in AGN evolution}
\author[S.I. Raimundo \& A.C. Fabian]{S.I. Raimundo$^{}$\thanks{E-mail: 
sijr@ast.cam.ac.uk} and A.C. Fabian\\
Institute of Astronomy, Madingley Road, Cambridge CB3 0HA}

\maketitle
\begin{abstract} 
The cosmological evolution of Active Galactic Nuclei (AGN) is important for understanding the mechanism of accretion onto supermassive black holes, and the related evolution of the host galaxy.
In this work, we include objects with very low Eddington ratio ($10^{-3} - 10^{-2}$) in an evolution scenario, and compare the results with the observed local distribution of black holes. We test several possibilities for the AGN population, considering obscuration and dependence with luminosity, and investigate the role of the Eddington ratio $\lambda$ and radiative accretion efficiency $\epsilon$ on the shape of the evolved mass function.
We find that three distinct populations of AGN can evolve with a wider parameter range than is usually considered, and still be consistent with the local mass function. In general, the black holes in our solutions are spinning rapidly. Taking fixed values for $\epsilon$ and $\lambda$ neither provides a full knowledge of the evolution mechanism nor is consistent with the existence of low Eddington ratio objects.
\end{abstract} 


\section{Introduction}
Understanding how Active Galactic Nuclei (AGN) work has been a major goal since the discovery of quasars \citep{schmidt63}. It is now believed that mass accretion onto a supermassive black hole (SMBH) is the mechanism that powers AGN (\citealt{salpeter64, lynden-bell69, rees84}), and this hypothesis is supported by recent evidence for massive black holes in the centre of most nearby galaxies (\citealt{kormendy&richstone95,richstone98}). The mass of the black hole (BH) is related to the properties of the host galaxy (\citealt{magorrian98, marconi&hunt03}), implying a close relation between the evolution of galaxies and the BH in their centres.

Soltan \citeyearpar{soltan82}, developed a method to obtain the total mass density of SMBH, $\rho_{\rm BH}$, from the emitted light and number counts of AGN, providing a way to trace the accretion history of these BHs, from the AGN luminosity function as a function of redshift. Following this approach, there have been attempts to compute the evolution of the BH mass function from a starting redshift, and compare it with the BH distribution in nearby galaxies, known as the local distribution (\citealt{small&blandford92, salucci99, marconi04, merloni04}; \citealt*{shankar07}). Progress has been made by the use of semi-analytical models and increasingly better observational results, although it is still hard to constrain the important factors affecting this calculation: the luminosity function at each redshift, the bolometric correction, the obscuration and the free parameters related with the accretion process.

There is evidence from the X-ray background that obscuration plays an important role in AGN evolution. The background is due to the integrated emission from AGN with high intrinsic absorption (e.g. \citealt*{brandt&hasinger05}), and by modelling it, \cite*{fabian&iwasawa99} found that most of the accretion onto AGN is obscured. One of the main goals of AGN studies is to get unbiased samples of objects, and for this, X-ray surveys reveal more AGN than found in other wavelength bands. The observations in the hard X-ray band (2-10 keV) are the least biased against obscuration, and therefore more suitable to observe and study AGN. \citealt{ueda03} (hereafter U03) show that the hard X-ray luminosity function is best-fitted by a luminosity dependent density evolution (LDDE), and that the number density of low luminosity AGN peaks at lower redshift than the high luminosity ones. To be able to convert the X-ray to total energy output, and obtain a knowledge of the multi-wavelength properties, a bolometric correction is needed, this is a function of source luminosity (e.g., \citealt{marconi04}), or more likely, the Eddington ratio (\citealt{vasudevan&fabian07}).\\
\indent The accretion mechanism is another important factor. There are two important free parameters to describe it; one, the radiative accretion efficiency $\epsilon$, representing the fraction of accreted mass radiated, establishes the relation between the bolometric luminosity and the growth rate of the BH, the other, the Eddington ratio $\lambda$, relates the AGN bolometric luminosity with the Eddington luminosity: $L=\lambda L_{\rm E}$, with $L_{\rm E}   = 4\pi cGMm_{\rm H}/\sigma_{\rm T}$. In previous work, the Eddington ratio has been taken to be either a constant, changing as a function of redshift, or a function of mass (see \citealt{shankar07} and references therein).\\
\indent Another factor is the hydrogen column density ($N_{\rm H}$) distribution, often called the $N_{\rm H}$ function. Several studies tried to parametrise it, U03 showed that the absorption is luminosity dependent, with $N_{\rm H}$ a decreasing function of the intrinsic luminosity, and found no significant dependence on redshift, although this question is still in debate (see for example \citealt{tozzi06}; \citealt*{treister&urry06}). 

To constrain the parameters involved in the growth of AGN, \cite{marconi04} compare the mass function from AGN relics with the local mass function, using a single AGN population. They find an acceptable match for $0.1 \leq \lambda \leq 1.7$, and $0.04 \leq \epsilon\leq 0.16$, with the values $\epsilon = 0.08$ and $\lambda = 0.5$ giving the minimum deviation between the two mass functions. However, recent results showed that the Eddington ratio can be significantly lower, with many objects having $\lambda\sim 10^{-2}-10^{-3}$, (\citealt{babic07}; \citealt*{fabian&vasudevan08}), at least one order of magnitude less than the \cite{marconi04} lowest acceptable value. This indicates that there is a wide spread in the range of $\lambda$, and that we have to consider low values for this parameter when computing the AGN mass function.

In this paper, we analyse how the free parameters $\lambda$ and $\epsilon$ influence the shape of the evolved AGN mass function, starting by considering one population of AGN, which is then split in two, according to obscuration. This leads to an implausible excess of high mass objects so we add a luminosity dependence for AGN with low $\lambda$, and consider the evolution of {\it three} different populations. The real situation is undoubtedly more complex than our simple models, but we expect to have captured the essence of AGN evolution in the ($\epsilon$, $\lambda$) plane.

We adopt standard cosmological parameters of H$_0 = 70$ km s$^{-1}$ Mpc$^{-1}$, $\Omega_{\rm m} = 0.3$ and $\Omega_{\rm \Lambda} = 0.7$.
\section{The black hole mass function}
To obtain a black hole mass function for $z = 0$, and compare it with the local observed mass function, we follow the procedure of \citet {marconi04} (hereafter M04) and \cite{shankar07}. Starting with an AGN luminosity function (LF) at certain redshift $z_{\rm i}$, and considering that all BHs were active, we follow their evolution with cosmic time, assuming that their growth and energy output are due to mass accretion only.

Using Soltan's argument (\citeyear{soltan82}), we can establish a relation between the total energy output, given by the integrated luminosity function, and the density of BH:
\begin{eqnarray} \label{soltan}
\rho_{\rm BH} = \frac {1-\epsilon}{\epsilon c^{2}\ln10}\int_0^{z_{\rm i}}\frac{dt}{dz}dz\int_{L_1}^{L_2}\frac{\partial \phi(L,z)}{\partial \log L} dL.
\nonumber
\end{eqnarray}
Here, $L$ is the bolometric luminosity, $\epsilon$ the radiative accretion efficiency and: 
\begin{eqnarray} \label{cosmology}
\frac{dt}{dz} = - \left[(1+z)H_0\sqrt{(1+z)^3\Omega_{\rm m} + \Omega_{\rm \lambda}} \right]^{-1},
\nonumber
\end{eqnarray}
for the cosmological model used in this work.
We are interested in the AGN distribution of masses, i.e., what is the number of objects with a certain mass $M$, per unit of comoving volume, and how it varies with redshift. For this, we use the continuity equation for the non-merging case, (see \citealt*{cavaliere71}; \citealt {small&blandford92}). Considering that the BH growth is only due to mass accretion, we assume that the total number of BHs is constant with time, and obtain:
\begin{eqnarray} \label{continuity}
\frac{\partial}{\partial t}\frac{\partial\phi(M,t)}{\partial M } + \frac{\partial}{\partial M}\left[\frac{\partial\phi(M,t)}{\partial M}\langle\dot{M}(M,t)\rangle\right] = 0.
\end{eqnarray}
In the case of merging, the right side of the continuity equation would not be zero, but equal to a source function, implying that the total number of BHs is not conserved.
In Eq. \ref{continuity}, $\phi(M,t)$ is the number of BH with mass $M$ per unit comoving volume for a certain cosmic time $t$ and $\langle\dot{M}(M,t)\rangle$ is the mean accretion rate for BH of mass $M$ (active and inactive):
\begin{eqnarray}\label{duty}
\langle \dot{M}(M,t)\rangle = \delta(M,t)\dot{M}(M,t).
\end{eqnarray}
Here $\delta(M,t)$ is the fraction of active (accreting) BH as a function of $M$ and $t$, called the duty cycle. $\dot{M}(M,t)$ is the accretion rate (or growth rate) of a BH with mass $M$ at time $t$.

The growth rate is a result of the mass that falls to the BH but is not converted into energy (see \citealt{marconi04}). If we have a total mass falling ($\dot{M}_{\rm tot}$), part of it is converted into energy $L=\epsilon\dot{M}_{\rm tot}c^2$, and the rest will accrete onto the BH: $\dot{M}=(1-\epsilon)\dot{M}_{\rm tot}$. Consequently,
\begin{eqnarray}
L = \frac{\epsilon}{(1-\epsilon)}c^2\dot{M}(M,t).
\nonumber
\end{eqnarray}
The energy output at a certain $t$ is due to active BHs, so
\begin{eqnarray}\label{phiLphiM}
\frac{d\phi(L,t)}{d\log L}d\log L = \delta(M,t)\frac{d\phi(M,t)}{dM}dM.
\end{eqnarray}
Using Eq. \ref{duty} and Eq. \ref{phiLphiM} to eliminate $\delta(M,t)$, and replacing in the continuity equation (\ref{continuity})
\begin{eqnarray}
\frac{\partial}{\partial t}\frac{\partial\phi(M,t)}{\partial M} +\frac{\partial}{\partial M}\left[\frac{d\phi (L,t)}{d\log L}\frac{d\log L}{dM}\dot{M}(M,t)\right] = 0. 
\nonumber
\end{eqnarray}
$L$ and $M$ are related by $L = \frac{\lambda c^2}{t_{\rm E}}M$, with $t_{\rm E} = \frac{\sigma_{\rm T}c}{4\pi Gm_{\rm H}}$, and assuming constant Eddington ratio ($\lambda$) and $\epsilon$, one can solve the continuity equation to obtain the mass function at a redshift $z_{\rm f}$:
\begin{eqnarray}
\lefteqn{\frac{d\phi(M,z_f)}{dM} = \frac{(1-\epsilon)\lambda^{2} c^{2}}{\epsilon t^{2}_{\rm E} \ln10}\int_{z_{\rm f}}^{z_{\rm i}}\frac{\partial}{\partial L}\left[\frac{d\phi(L,z)}{d\log L}\right] \left|\frac{dt}{dz}\right| dz}
\end{eqnarray}
This equation tells us that the mass function at a certain redshift, can be obtained with a Soltan-type argument, by tracking the energy that was produced due to mass accretion from $z_{\rm i}$ to $z_{\rm f}$. We assume that the initial mass function (at redshift 3), does not contribute significantly to the final mass function, since the major black hole growth occurs for $z < 3$ (M04).

To be able to compare this evolved mass function with the local one, the previous equation has to be evaluated for $z_{\rm f} = 0$. We have to assume a certain luminosity function, a bolometric correction, set the initial conditions and give values to $\epsilon$ and $\lambda$.
For the LF, we consider the luminosity-dependent density evolution (LDDE) of U03, which describes $\frac{\partial \phi (L_{\rm X},z)}{\partial \log L_{\rm X}}$, the number of AGN with intrinsic luminosity $L_{\rm X}$ per unit of comoving volume per $\log L_{\rm X}$, as function of the luminosity and the redshift:
\begin{eqnarray}
\frac{d\phi(L_{\rm X},z)}{d\log L_{\rm X}}=A\left[\left(\frac{L_{\rm X}}{L_{\rm \star}}\right)^{\rm \gamma_1}+\left(\frac{L_{\rm X}}{L_{\rm \star}}\right)^{\rm \gamma_2}\right]^{-1}e(z,L_{\rm X}),
\nonumber
\end{eqnarray}
where $e(z,L_{\rm X})$ is the evolution term,
\begin{eqnarray}
e(z,L_{\rm X})=
\begin{cases}
(1+z)^{\rm p_1} & z<z_{\rm c}(L_{\rm X}), \\
[1+z_{\rm c}(L_{\rm X})]^{\rm p_1-p_2}(1+z)^{\rm p_2} & z \geq z_{\rm c}(L_{\rm X}),
\end{cases}
\nonumber
\end{eqnarray}
and the redshift cut off,
\begin{eqnarray}
z_{\rm c}(L_{\rm X})=
\begin{cases}
z_{\rm c}^{\star} & L_{\rm X} \geq L_{\rm a}, \\
z_{\rm c}^{\star}\left( \frac {L_{\rm X}}{L_{\rm a}}\right)^{\rm \alpha} & L_{\rm X}<L_{\rm a}.
\end{cases}
\nonumber
\end{eqnarray}
The parameters have the values: A = $5.04\times10^{-6}$ Mpc$^{-3}$; L$_{\rm a}$ = 10$^{44.6}$ erg s$^{-1}$; L$_{\star} = 10^{43.94}$ erg s$^{-1}$; $\gamma_{1}$ = 0.86; $\gamma_{2}$ = 2.23; p$_1$ = 4.23; p$_2$ = -1.5; z$_c^{\star}$ = 1.9; $\alpha$ = 0.335. 
This luminosity function has a discontinuity in its derivative, which leads to unphysical solutions for the mass function. To avoid that, we rewrite the luminosity function to make it continuous in its derivatives (A. Marconi, private communication):
\begin{eqnarray}
\frac{d\phi(L_X,t)}{d\log L_X} = \left[\left(\frac{d\phi(L_X \geq L_a,t)}{d\log L_X}\right)^{-\beta}+\left(\frac{d\phi(L_X < L_a,t)}{d\log L_X}\right)^{-\beta}\right]^{-\frac{1}{\beta}}
\nonumber
\end{eqnarray}
with $\beta=8$.

For the bolometric luminosity $L$, we use the Eddington ratio dependent bolometric correction $\kappa (\lambda) = L/L_{\rm X}$ (\citealt{vasudevan&fabian07}), with $\kappa (\lambda) = 19.3$ for $\lambda \leq 0.1$,  $\kappa (\lambda) = 54.5$ for $\lambda \geq 0.3$, and an intermediate value of $\kappa (\lambda) = 36.9$ for the other Eddington ratios. To account for Compton-thick AGN, we multiply our mass function by a constant factor of 1.6, following the result by \cite{risaliti99}.
In this work we consider $z_{\rm i} = 3$, and that all BHs were active at that time ($\delta(M,z_{\rm i})=1$). Using
\begin{eqnarray}
\frac{d\phi (L,z)}{d\log L}d\log L = \frac{d\phi (L_{\rm X}, z)}{d \log L_{\rm X}}d\log L_{\rm X},
\nonumber
\end{eqnarray}
and since $\kappa (\lambda)$ is constant for an assumed value of $\lambda$, we finally obtain
\begin{eqnarray}\label{massf}
\frac{d\phi(M,0)}{dM}M=\frac{(1-\epsilon)\lambda L}{\epsilon t_{\rm E}\ln 10}\int_{0}^{3}\frac{\partial}{\partial L}\left[\frac{d\phi(L_{\rm X},z)}{d\log L_{\rm X}}\right] \times\left|\frac{dt}{dz}\right| dz.
\end{eqnarray}\\
For a simple model of one population of AGN, we apply the method described before, noting that each set of parameters $\lambda$ and $\epsilon$ give a different mass function for the same luminosity. To add the effect of obscuration, and understand the role of the absorbed AGN in the shape of the mass function, we consider the AGN population to be separated in two groups with different absorption column densities $N_{\rm H}$: unabsorbed ($10^{20}$cm$^{-2}\leq N_{\rm H} <10^{22}$cm$^{-2}$), and absorbed ($10^{22}$cm$^{-2}\leq N_{\rm H} <10^{24}$cm$^{-2}$). We define  
\begin{eqnarray}\label{Nh}
\mathcal{N}_{\rm H}(L_{\rm X}) = \int_{N_1}^{N_2} f(L_{\rm X};N_{\rm H})d\log N_{\rm H},
\end{eqnarray}
where $f(L_{\rm X};N_{\rm H})$ is the probability distribution function from U03, normalised to unity in the Compton-thin region $20.0 \leq \log N_{\rm H} \leq 24.0$. We find $\mathcal{N}_{\rm H}^{\rm abs}(L_{\rm X})$ integrating Eq. \ref{Nh} between the values of column density assumed for absorbed AGN, and  $\mathcal{N}_{\rm H}^{\rm un}(L_{\rm X})$ integrating in a similar way for the unabsorbed ones. That is, for each value of $L_{\rm X}$, we split our population in two groups, with constant number fractions given by the $\mathcal{N}_{\rm H}(L_{\rm X})$ function. 
In summary we have:
\begin{eqnarray}\label{Nha}
\frac{d\phi (M,0)}{dM}^{\rm abs}=\mathcal{N}_{\rm H}^{\rm abs}(L_{\rm X})\times \frac{d\phi (M,0)}{dM},
\end{eqnarray}
\begin{eqnarray}\label{Nhu}
\frac{d\phi (M,0)}{dM}^{\rm un}=\mathcal{N}_{\rm H}^{\rm un}(L_{\rm X})\times \frac{d\phi (M,0)}{dM}.
\end{eqnarray}
\section{Constraints on the free parameters}
\begin{figure}
\begin{centering}
\includegraphics[width=0.8\columnwidth]{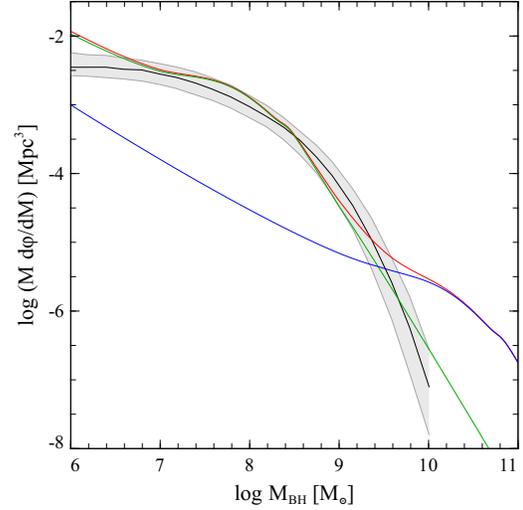}
\caption {Comparison between the evolved number of AGN per unit volume and the local mass function (in black) by \citet{marconi04}. The grey area is the 1$\sigma$ error in the M04 local mass function. The green curve represents the unabsorbed AGN, and the blue curve the absorbed AGN. The red curve is the total population (absorbed plus unabsorbed). The values used were $\lambda_{\rm abs}=10^{-3}$, $\epsilon_{\rm abs}=0.085$, $\lambda_{\rm un}=0.56 $, $\epsilon_{\rm un}=0.21$ with the \citet{ueda03} $N_{\rm H}$ distribution. For a low value of $\lambda$, a two population model predicts a higher number of high mass black holes than that observed.}
\end{centering}
\end{figure}
To constrain the free parameters $\lambda$ and $\epsilon$, we compare our evolved AGN mass function with the local mass function. As a reference, we use the local mass function in M04 and the respective 1$\sigma$ error, in the range $(10^6-10^{10}) $M$_{\rm \odot}$ which was obtained using both the $M_{\rm BH} - \sigma_{\rm \ast}$ (velocity dispersion), and the $M_{\rm BH} - L_{\rm bul}$ (bulge luminosity) relations.
The method to obtain the mass function is as follows: for each mass value, we compute the corresponding $L_{\rm X}$ by first using $L = \lambda L_{\rm E}$ to obtain the bolometric luminosity from the mass, and then $\kappa(\lambda)$ to convert from bolometric to intrinsic luminosity. We then replace the U03 luminosity function and the bolometric correction in Eq. \ref{massf}, differentiate in $L$ and integrate from $z = 0$ to $z = 3$ to obtain $\frac{d\phi (M,0)}{dM}$. 
For the case of two or more populations we use Eq. \ref{Nh} to compute the $\mathcal{N}_{\rm H}$ function for each value of $L_{\rm X}$, and replace it in Eq. \ref{Nha} for the absorbed and Eq. \ref{Nhu} for the unabsorbed, allowing each population to have a different Eddington ratio and accretion efficiency. 

For one population, with only two free parameters, we find no value of $\epsilon$ or $\lambda$ for which the evolved mass function is consistent with the local one within a 1$\sigma$ error. Considering a black hole mass range of ($10^{7} - 10^{9}$)$M_{\odot}$, we obtain agreement but only for high Eddington ratios ($\lambda > 0.1$). To search for alternative scenarios and be able to have lower $\lambda$ we study the effect of obscuration, by considering the $N_{\rm H}$ function by U03.

\subsection{Two populations}

We start by assuming two populations, one of obscured and one of unobscured AGN. The agreement between the local and the evolved number density of AGN can be obtained by changing simultaneously the free parameters for the absorbed and unabsorbed objects. To constrain the allowed ranges for the accretion efficiency and the Eddington ratio, we search parameter space, find the evolved number density of AGN for every combination of the four parameters ($\lambda_{\rm abs}, \epsilon_{\rm abs}, \lambda_{\rm un}, \epsilon_{\rm un}$), and compare it with the local number density from M04.
We search the parameter range: $10^{-3} \leq \lambda \leq 1$,  $10^{-2} \leq \epsilon \leq 0.42$. The upper value of $\epsilon$ is limited by the maximum accretion onto a Kerr black hole, if we consider a case of a radiatively efficient system, the radiative accretion efficiency $\epsilon$ is equal to the mass to energy efficiency ($\eta$). In a more general case, $\epsilon \leq \eta$.

With two populations, we do not find parameters that result in a mass function within 1$\sigma$ of the local mass function. We only find good agreement if we limit the mass range to $M_{\rm BH} = (10^{7} - 10^{9})$M$_{\rm \odot}$, but for low values of $\lambda$, it predicts a higher density of high mass black holes ($>10^{9}M_{\rm \odot}$) than that observed, which is not consistent. In Fig.~1 we chose a set of values to illustrate our problem, the total mass function in red is compared with the local mass function from M04. The blue and green lines represent the absorbed and unabsorbed populations respectively, with fixed parameters. For low values of $\lambda$, although it is possible to obtain good agreement in a limited mass range, the sum of the absorbed and unabsorbed populations (red line), predicts a higher number of high mass black holes than observed. This is a common problem, both for M04 luminosity dependent and \cite{vasudevan&fabian07} Eddington ratio dependent bolometric corrections. When $\lambda$ is lowered ($\lambda < 10^{-2}$), the mass function shifts down and right, which means that it raises the number density of high mass black holes. Nevertheless, low $\lambda$ objects are observed in the Universe, and we require that they agree with the local mass function.

\subsection{Three populations}
To obtain agreement in the high mass black hole range, the number of AGN with low Eddington ratio has to drop with luminosity. By hypothesis, and in agreement with the $N_{\rm H}$ function in U03, we assume that the low $\lambda$ objects are included in the obscured population. We therefore divide our absorbed population in two: the absorbed objects with lower Eddington ratio ($10^{-3}<\lambda<10^{-2}$), and objects with higher Eddington ratio ($10^{-2}<\lambda<1$). This way, our AGN population is divided into three groups: obscured objects with lower $\lambda$ and numbers dropping with increasing luminosity, obscured objects with higher $\lambda$, and unobscured objects ($10^{-2}<\lambda<1$).

For the luminosity dependence, we define a simple decay for the fraction $f(L_{\rm X})$, the number of objects with low $\lambda$ divided by the total number of absorbed objects for a certain luminosity:
\begin{eqnarray}
f(L_{\rm X}) = S\log(L_{\rm X}) + b,
\nonumber
\end{eqnarray}
where $S$ is the free parameter, and $b$ is determined by assuming that for $L_{\rm X} < 10^{41}$erg s$^{-1}$ all the absorbed AGN have low $\lambda$ ($f = 1$). Therefore, the mass functions for the three populations are: \\
- Absorbed with low $\lambda$:
\begin{eqnarray} 
\frac{d\phi (M,0)}{dM}^{\rm labs}=f(L_{\rm X})\times\mathcal{N}_{\rm H}^{\rm abs}(L_{\rm X})\times \frac{d\phi (M,0)}{dM}
\end{eqnarray}
- Absorbed:
\begin{eqnarray} \frac{d\phi (M,0)}{dM}^{\rm abs}=(1-f(L_{\rm X}))\times\mathcal{N}_{\rm H}^{\rm abs}(L_{\rm X})\times \frac{d\phi (M,0)}{dM}\end{eqnarray}
- Unabsorbed:
\begin{eqnarray} \frac{d\phi (M,0)}{dM}^{\rm un}=\mathcal{N}_{\rm H}^{\rm un}(L_{\rm X})\times \frac{d\phi (M,0)}{dM}\end{eqnarray}
In terms of free parameters, we now have ($\lambda_{\rm abs}, \epsilon_{\rm abs}, \lambda_{\rm un}, \epsilon_{\rm un}, \lambda_{\rm labs}, \epsilon_{\rm labs}, S$).
\begin{figure}
\begin{centering}
\includegraphics[width=0.9\columnwidth]{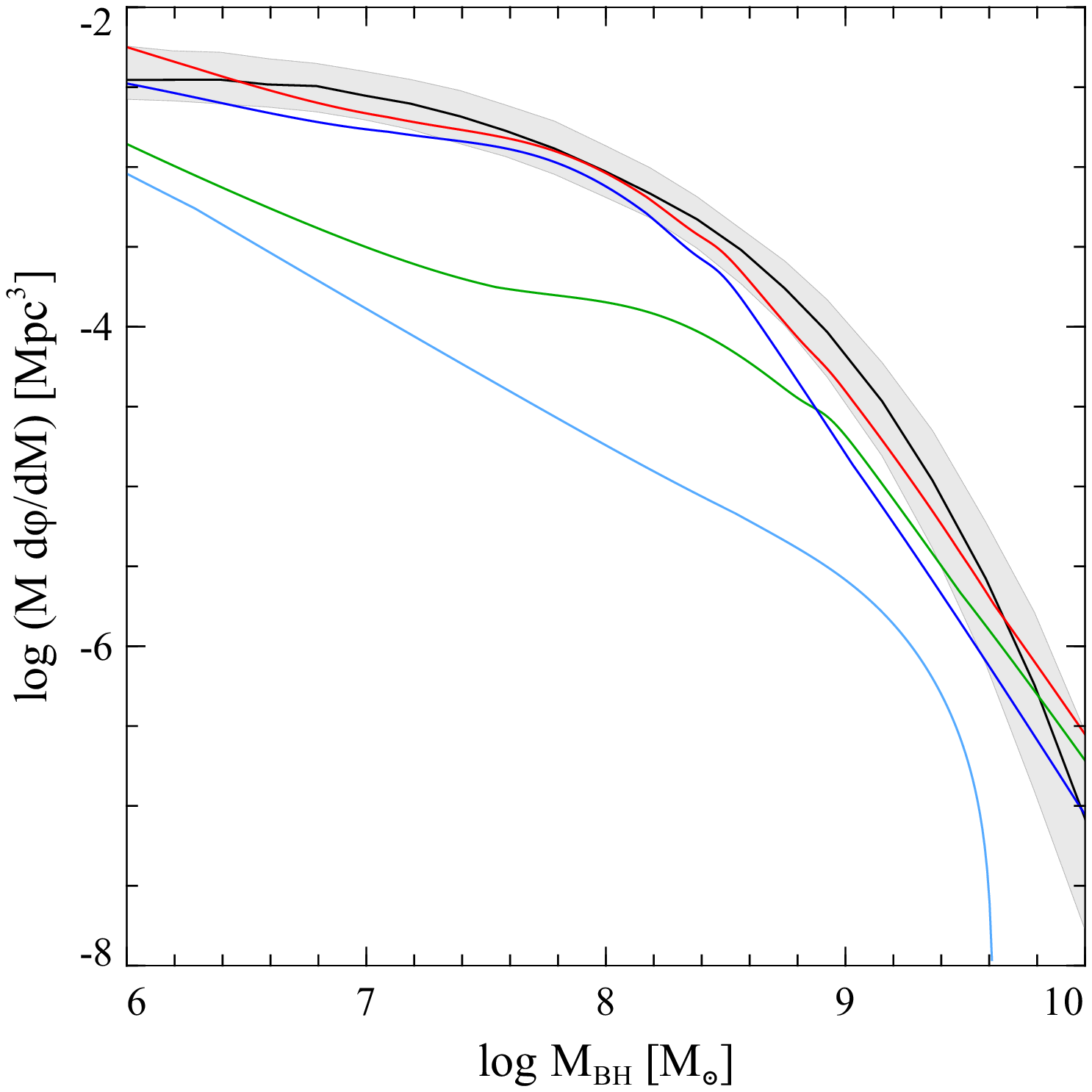}
\caption {Comparison between the evolved number of AGN per unit volume and the local mass function (in black) by \citet{marconi04}. The grey area is the 1$\sigma$ error in the M04 local mass function. The green curve represents the unabsorbed AGN, and the blue curves the absorbed AGN: light blue for the absorbed objects with low $\lambda$ and dark blue for the others. 
The red curve is the total population (absorbed, absorbed low $\lambda$ and unabsorbed). The values used were $\lambda_{\rm labs}=0.007$, $\epsilon_{\rm labs}=0.21$, $\lambda_{\rm abs}=0.5$, $\epsilon_{\rm abs}=0.068$, $\lambda_{\rm un}=0.079 $, $\epsilon_{\rm un}=0.21$ and $S = -0.3$ with the \citet{ueda03} $N_{\rm H}$ distribution.}
\end{centering}
\end{figure}
Following this method, we find several sets of parameters for which there is good agreement with the local mass function. In Fig.~2 we have an example of a set of parameters for which there is good agreement, the grey area is the 1$\sigma$ error of the M04 local mass function (in black) and the red line is our total evolved mass function. The green line represents the unabsorbed population, and the blue lines the absorbed: light blue for the low $\lambda$ and dark blue for the others. As we can see, a decay in $f$ with luminosity allows the low $\lambda$ population to agree in the high black hole mass range, and we obtain very good agreement between our evolved mass function and the local one. 
In terms of the effect of a parameter change, lowering $\epsilon$, makes the mass growth of the BHs more efficient, and the curves in Fig.~2 go up. $\lambda$ shifts the curves to the high mass or low mass end. An increase in $\lambda$ corresponds to a shift to the left and up, a decrease makes the curves go right and down.
The plot in Fig.~3 shows the result of our search in the parameter range described for the two populations and also with $10^{-3}<\lambda_{\rm labs}<10^{-2}$, $10^{-2}<\epsilon_{\rm labs}<0.42$ and $-1 < S < 0$. Each population is described by a set of three points in the plot: light blue (light grey), dark blue (black) and green (medium grey). The sum of these three populations gives the total one, which is evolved and compared with the local mass function. We plot all the values of $\lambda$ and $\epsilon$ for which the evolved mass function is within a 1$\sigma$ error from the local one.
Note that the number of points are not the same for all populations because an unabsorbed population can have, for example, more than one absorbed or low lambda absorbed population for which the mass function agrees well with the local one.

In terms of the general features, we see that we obtain a wide range of possible parameters. Each population is described by a point in Fig.~3, which means that the total mass function (red line on the previous plots), is described by a definite set of three points, light blue (light grey), dark blue (black) and green (medium grey), and not a random combination. The absorbed and unabsorbed, blue (light grey and black) and green (black), populations, have values of $\lambda$ that are in general higher that $0.1$, and the absorbed with low $\lambda$, light blue (light grey), spread in the whole parameter range ($10^{-3}-10^{-2}$). We notice a void, due to the fact that for the absorbed with high $\lambda$ and unabsorbed, an Eddington ratio $ \gtrsim 0.1$ causes the evolved mass function to disagree from the local mass function in the high mass range. Allowing the Eddington ratio for the low $\lambda$ absorbed objects to go higher than $10^{-2}$ would fill in the void. Values for the slope $S$, are found between $-0.75$ and $-0.25$, which means that the values for the luminosity $L_{\rm X}$ where $f$ reaches zero range from $10^{42.3}$erg s$^{-1}$ to $10^{45}$erg s$^{-1}$. These values correspond to a steeper decay than for example the absorbed fractions from \cite{hasinger08} or \cite{treister08}, which combine absorbed objects of all $\lambda$.
The values for the accretion efficiency are above $0.035$, and in this scenario, lower for absorbed objects than unabsorbed objects in the same range of $\lambda$. We do not find evidence for advection dominated accretion flows (ADAF) behaviour in our low efficiency objects.
As we can see from Fig~3, almost all possible parameters are above $\epsilon = 0.057$, which is the efficiency corresponding to a Schwarzschild (non-rotating) black hole, indicating that the black holes must be spinning. In fact, most of the black holes in our solutions are rapidly spinning. A population of low $\lambda$ objects with low efficiency, light blue (light grey) points, or low efficiency absorbed objects, dark blue (black) points, can only represent the local mass function if combined with unabsorbed, green (medium grey), objects with high efficiencies ($\epsilon > 0.1$).

We have assumed that Compton-thick objects contribute to the mass function equally in all the mass range, and set a correction factor of 1.6 to account for them. A change in this value would not affect qualitatively the shape of the relation between $\epsilon$ and $\lambda$ showed in Fig.~3.
To evaluate the errors, we used two approaches: a higher multiplying correction factor and a similar method to \cite{gilli07}. In terms of the mass functions, a higher multiplying factor shifts all the curves up, making the lower values of $\epsilon$ in Fig.~3 be inconsistent with the local mass function. In summary, a higher fraction of Compton-thick objects would imply that the local mass function can be reproduced using higher accretion efficiencies than the ones from Fig.~3. For the \cite{gilli07} $N_{\rm H}$ function, we obtain results that do not differ significantly from the ones with the U03 $N_{\rm H}$ function, the main difference is that the parameter values for the absorbed objects (high $\lambda$), are found in the same range as the unabsorbed, that is, higher than the ones from U03.

\section{Summary and Conclusions}
\begin{figure}
\includegraphics[width=0.9\columnwidth,clip=true]{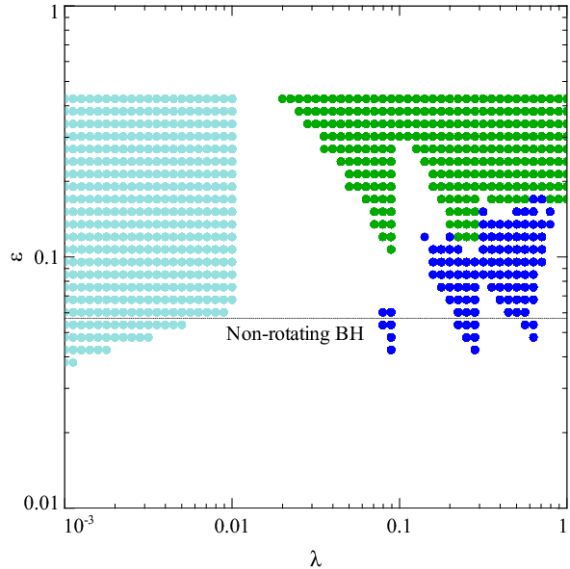}
\caption {Allowed values of Eddington ratios ($\lambda$) and radiative accretion efficiencies ($\epsilon$) for absorbed, blue (light grey and black) and unabsorbed, green (medium grey) AGN. Light blue points (light grey) represent the fraction of absorbed objects that have low Eddington ratios $10^{-3} < \lambda < 10^{-2}$, and dark blue (black) the others $10^{-2} < \lambda < 1$. Unabsorbed objects with $10^{-2} < \lambda < 1$ are plotted in green (medium grey). Each evolved mass function is characterised by a fixed set of three points from the plot.}
\end{figure}
We have tried to reconcile the existence of low Eddington ratio AGN, with the local mass function. Our approach has been to try the simplest solution and then add features to obtain good agreement. In order to include a population of low $\lambda$ objects and a large spread in possible $\lambda$ values, we find that at least three populations are required, with different absorption properties depending on intrinsic luminosity for the lowest $\lambda$ objects. AGN evolution is not a straightforward problem, and may in future require more complex behaviour to be considered.
We find that most of the black holes in our solutions have high accretion efficiencies ($\epsilon > 0.1$), and therefore must be spinning rapidly, which has implications for the merger and accretion histories (\citealt{volonteri05}; \citealt*{king&pringle06}).

\section{Acknowledgements}
We thank the referee for helpful comments and A. Marconi for providing useful details about his work. SR acknowledges financial support from FCT - Funda\c c\~ao para a Ci\^encia e a Tecnologia (Portugal). ACF thanks The Royal Society for support. 

\bibliographystyle{mnras}
\bibliography{AGN}

\end{document}